\documentclass[useAMS,usenatbib]{mn2e}
\usepackage{psfig}
\usepackage{natbib}

\def\ltsima{$\; \buildrel < \over \sim \;$}
\def\lsim{\lower.5ex\hbox{\ltsima}}
\def\gtsima{$\; \buildrel > \over \sim \;$}
\def\gsim{\lower.5ex\hbox{\gtsima}}
\def\be{\begin{equation}}
\def\ee{\end{equation}}
\def\no{\noindent}
\def\Mesz{M\'esz\'aros}

\newcommand{\apj}{ApJ}
\newcommand{\apjl}{ApJL}
\newcommand{\apjs}{ApJS}
\newcommand{\mnras}{MNRAS}

\newcommand{\aap}{A\&A}

\begin{document}

\title[ ``Orphan'' afterglows in the USJ model] {``Orphan'' afterglows
in the Universal Structured Jet Model for $\gamma$-ray bursts}

\author[Rossi, Perna \& Daigne]{Elena M. Rossi $^{1,2}$,  Rosalba Perna$^{1,3}$ \& Fr\'{e}d\'{e}ric Daigne$^{4,5}$ \\
$^1$JILA, University of Colorado at Boulder, 440 UCB
Boulder, CO 80309-0440 \\ 
$^2$ Chandra Fellow \\
$^3$ Department of Astrophysical and Planetary Sciences, University of Colorado \\
$^4$ Institut d'Astrophysique de Paris, UMR 7095 CNRS --
Universit\'{e} Pierre et Marie Curie-Paris VI, 98 bd Arago, 75014
Paris, France.\\
$^5$ Institut Universitaire de France \\
\tt e-mail: emr@jilau1.colorado.edu (EMR), rosalba@jilau1.colorado.edu (RP) daigne@iap.fr (FD)}

\maketitle

\begin{abstract}
The paucity of reliable achromatic breaks in Gamma-Ray Burst afterglow
light curves motivates independent measurements of the jet aperture. 
 Serendipitous searches of afterglows, especially at radio wavelengths, have
long been the classic alternative. These survey data have been interpreted assuming a
uniformly emitting jet with sharp edges (``top-hat'' jet), in which case
the ratio of weakly relativistically beamed afterglows to GRBs scales with the jet
solid angle. In this paper, we consider, instead,  a very wide outflow with a luminosity
that decreases across the emitting surface.  
 In particular, we
adopt the  universal structured jet (USJ) model,   that
is an alternative to the top-hat model for the structure of the jet.
 However, the interpretation of the survey data is very
different: in the USJ model we only observe the emission within
 the jet aperture and the observed ratio of prompt emission rate to
 afterglow rate should solely depend on selection effects.
We compute the number and rate
of afterglows expected in all-sky snapshot observations as a
function of the survey sensitivity. We find that the current
(negative) results for OA searches are in agreement with our
expectations. In radio and X-ray bands this was mainly due to the low
sensitivity of the surveys, while in the optical band the sky-coverage
was not sufficient. 
In general we find that X-ray surveys are poor tools for OA
searches, if the jet is structured. On the other hand, 
the FIRST radio survey and future instruments
like the Allen Telescope Array (in the radio band) 
and especially GAIA, Pan-Starrs  and LSST (in the optical band) 
will have chances to detect afterglows.

\end{abstract}

\medskip

\section{Introduction}
\label{sec:intro}

Surveys for transient sources may detect Gamma-Ray Burst (GRB)
afterglows. In this paper, we call an ``orphan'' afterglow {\it any}
afterglow associated with such serendipitous searches, as opposed 
to ``triggered'' GRB afterglows, localized through the preceding
prompt $\gamma$-ray emission. Rhodes (1997) 
suggested that these surveys could be used 
to put constraints on the geometrical
beaming angle of the GRB jets in the ``top-hat'' (thereafter TH)
model.  In this model, GRBs are assumed to be uniformly emitting
within a cone of angle $\theta_{\rm jet}$ with sharp edges, where the
luminosity drops suddenly to an undetectable level.

The suggestion by Rhodes (1997) is based on the fact that the prompt
$\gamma$-ray emission is relativistically beamed within an angle
$\theta_{\rm jet}+1/\Gamma$, where $1/\Gamma << \theta_{\rm jet} $.  Thus,
if the line of sight lies outside this angle, the GRB is
unlikely to be detected. However, as the outflow slows down in the afterglow
phase, the visible region increases to eventually encompass the
observer's line of sight.  At late times, $\Gamma\sim 1$ and the
emission is roughly isotropic. This behaviour would suggest that
more long-wavelength transients than
 $\gamma$-ray ones {\rm may} be expected.  
 Detections of transients at different wavelengths 
may thus be used to constrain the beaming factor $b
\propto \theta_{\rm jet}^{-2}$.
A measure of this quantity would be of
great importance as it would allow one to calibrate $\theta_{\rm jet}$
and then estimate the true GRB rates ($\propto b$) and energetics
($\propto b^{-1}$).

Searches for OAs have been performed at various
wavelengths, but none of them has yielded a firm detection. 
Several authors have used observations in the radio band to
constrain the GRB rate and energetics (e.g. Perna \& Loeb 1998; Woods \&
Loeb 1999; Paczyski 2001; Levinson 2002; Gal-Yam et al. 2006), since
late time (i.e. nearly isotropic) afterglow emission peaks in this
band.  The simplest and most common assumption is that the predicted
number of OAs in a snapshot observation is {\it proportional} to the
beaming factor. Perna \& Loeb (1998) used the lack of detections to
set an upper limit of $b\la 10^3$.  More recently, Levinson et
al. (2002) and Gal-Yam et al. (2006) compared their data with a more
detailed model for radio afterglows.  They showed that the number of
expected OAs in a flux limited survey is {\it inversely} proportional
to $b$ and they place a lower limit of $b \gsim 60$.

X-ray survey data have been used to search for OAs by Grindlay (1999) 
and by Greiner et
al. (2000), while optical searches have been more numerous
(Schaefer et al. 2002; Vanden Berk et al. 2002; Becker et al. 2004;
Rykoff et al. 2005; Rau et al. 2006, Malacrino et al. 2007). Using
shorter wavelengths than radio to constrain the beaming factor
necessarily requires a careful comparison with theoretical predictions
(e.g. Totani \& Panaitescu 2002; Nakar, Piran \& Granot 2002), since
the emission is likely to be still relativistically beamed when OAs
are detected in those bands.  Malacrino et al. (2007, 2007b) put the
tightest optical constraints so far (see their fig.3). Their
non-detection resulted in an upper limit for the number of OAs on the
sky that is marginally consistent with the predictions of Totani \&
Panaitescu (2002, thereafter TP02) and consistent with Nakar et al. 
(2002, thereafter N02) and Zou et al. (2007, thereafter Z07).

 The purpose of this paper is to predict results for searches of
 afterglows in surveys, for a different GRB jet structure.
 Currently, OA surveys generally aim at constraining the jet angle.
 However, if GRB outflows are {\it not} geometrically beamed but they 
 rather have an anisotropic luminosity distribution, 
 the interpretation of data within the TH scenario would be misleading.

 Numerical
 simulations of collapsing massive stars (e.g. MacFadyen \& Woosley
 1999) show that the jet emerges from the star with an energy
 distribution $E(\theta)$ and Lorentz factor $\Gamma(\theta)$ that
 vary as a function of the angle $\theta$ from the jet axis. It has
 been shown (Rossi, Lazzati \& Rees 2002; Zhang \& \Mesz ~2002) that,
 if $E(\theta) \propto \theta^{-2}$ the diversity of afterglow light
 curves can be ascribed to different viewing angles within the context
 of an universal structured jet (USJ). In the USJ model, the outflow
 is geometrically wide.  In this paper, we will postulate that for
 each GRB two simultaneous and opposite jets with $\theta_{\rm
 jet}=90^{\circ}$ are produced. In this model $b=1$.
 The feature of having emission into
 $4\pi$ solid angle is attractive since it can explain the lack of OA
 detections, while the non-uniform energy distribution allows one to
 avoid the huge energy requirement, demanded by GRBs with an isotropic
 equivalent energy $\gsim 10^{54}$ ergs (e.g.~GRB 990123).
 An important consequence of the energy distribution law
 $E(\theta) \propto \theta^{-2}$ is that it establishes a unique
 relation between the viewing angle and the observed luminosity,
 once a radiation efficiency law with angle is assumed. 
 Thus, unlike in the TH model, the observed luminosity function is 
 not a free parameter. Consequently, the uncertainties in the 
 predicted OA rates in the USJ scenario are smaller than in the TH model (see \S~\ref{sec:USJvsUJ}).

In this paper, we use the USJ framework to compute the
expected number of transients in an all-sky snapshot  and
their rate as a function of the survey sensitivity. The aim is
pursued by means of Monte Carlo simulations of the afterglow
properties, as observed by X-ray, optical and radio surveys.  The
procedure is described in \S~\ref{sec:simule}. We show prospects for OA
detections with current and future surveys in \S~\ref{sec:results}. This allows us
to identify the best survey characteristics to increase the chance of
detection and single out the most promising future missions.
A comparison of our results with the top-hat predictions is performed in \S~\ref{sec:USJvsUJ}.
 Finally,  we discuss and conclude our work in \S~\ref{sec:fine}.

Throughout the paper, we assume a flat Universe with $H_{0}=73\ \mathrm{km ~s^{-1} Mpc^{-1}}$,
$\Omega_\mathrm{m}=0.3$ and $\Omega_\mathrm{\Lambda}=0.7$.

\section{Simulating the population of orphan  afterglows}
\label{sec:simule}

We use Monte Carlo methods to simulate the population of GRB orphan
afterglows in the USJ. GRBs are randomly generated on the sky with a
probability distribution in redshift that traces the star formation
rate (SFR) (\S~\ref{sec:sfr}). We use the external shock model to
compute the afterglow luminosity curve (\S~\ref{sec:code}). The
probability function for the viewing angle $\theta$ is given by the
fraction of the solid angle associated with that angle,
$P(\theta)\propto \sin(\theta)$.  Our simulation yields for radio,
optical and X-ray bands the distribution of afterglow fluxes and the
total number of afterglows on the sky for a snapshot
observation, together with the average time $T_{\rm th}$ that
an afterglow remains detectable in the sky, as a function of the
detection threshold. Finally, we compute the OA detection rate for any
flux limited survey where the observation time is much greater than
$T_{\rm th}$.

\subsection{Formation rate and $\gamma-$ray luminosity function}
\label{sec:sfr} 
We assume that the GRB population in the universe is described by a
redshift-independent luminosity function, a redshift
distribution and a distribution of the spectral parameters.

\no
In the USJ, the isotropic equivalent kinetic energy in the afterglow phase has the angular dependence
\be
E(\theta) = \frac{E_{\rm c}} {1+\left(\frac{\theta}{\theta_{\rm c}}\right)^{2}},
\label{eq:usj}
\ee

\no
where the bright central spine with angular size $\theta_{\rm c}$ 
has a maximum  kinetic energy $E_{\rm c}=E(0)$.
The expected luminosity function follows from 
$P_{\rm GRB}(L) = P(\theta) \frac{{\rm d} \theta}{{\rm d} L}$ (Rossi et al.~ 2002),
\begin{equation} 
 P_{\rm GRB}(L) \propto \frac{\sin{\left(\theta_{\rm c} \sqrt{\frac{L_{\rm c}}{L}-1}\right)}} {\sqrt{\frac{L_{\rm c}}{L}-1}} \,\,\left(\frac{L_{\rm c}}{L}\right)^2, 
\label{eq:lumfun}
\end{equation}
where $L$ [erg s$^{-1}$] is the isotropic equivalent bolometric
peak luminosity and $L_{\rm c}$ is proportional to $E_{\rm c}$ (see eq.~\ref{eq:ec}).
In equation~\ref{eq:lumfun}, we assume that the
$\gamma-$ray emission efficiency and the ratio of mean luminosity to
peak luminosity is independent of the angle.

\no
The GRB comoving rate $R_{\rm GRB}(z)$ [yr$^{-1}$ Mpc$^{-3}$] is assumed
to follow the comoving rate $ R_{\rm SN}(z)$ [yr$^{-1}$ Mpc$^{-3}$] of
Type II supernovae 
$
R_\mathrm{GRB}(z) = k \times R_\mathrm{SN}(z)
$, 
where $k\equiv R_\mathrm{GRB}/R_\mathrm{SN}$ is a free parameter of the model.
 We assume that these
arise from stars with masses above 8
$M_{\odot}$ and that the initial mass function has a Salpeter form (e.g. Porciani \&
Madau 2001). Daigne, Rossi $\&$ Mochkovitch (2006, thereafter DRM06)
found that the above prescription is dubious for redshift greater
than 2.  However, most of the OAs that are detected in a survey are
located at lower redshifts, where this assumption appears to hold. 
In our analysis, we find, in fact, that the mean OA
redshift for any reasonable flux threshold is never greater than $z=2$.
 The
star formation rate we adopt to derive $ R_{\rm SN}$ (dubbed SFR$_{\rm
2}$) and its comparison with data (Hopkins 2004) are shown in fig.~1 of
DRM06. This SFR saturates beyond $z \sim 2$ at a level of
0.2 M$_{\odot}$ yr$^{-1}$ Mpc$^{-3}$. Even though this behavior is
consistent with data at the one sigma level, the flat extrapolation of our
SFR after the peak remains questionable, since high-z data are
plagued by uncertainty on the amount of dust extinction. However, as discussed above,
we do not expect that uncertainties in the high-z
behaviour of $R_{\rm GRB}(z)$ appreciably affect our results. The
GRB redshift probability function is thus given by
\begin{equation}
p(z) \propto \frac{dV}{dz}\frac{R_\mathrm{GRB}(z)}{1+z}\ ,
\label{eq:probaz}
\end{equation}
where the comoving volume equals 
$$
\frac{dV}{dz}=\frac{c}{H_{0}} \frac{4\pi
D_\mathrm{L}^{2}(z)(1+z)^{-2}}{\sqrt{\Omega_\mathrm{m}\left(1+z\right)^{3}+\Omega_\mathrm{\Lambda}}}
$$
and $D_\mathrm{L}(z)$ is the standard luminosity distance.

\no
The spectral properties of GRBs are described by the
distribution of their peak energy $E_\mathrm{p}$ and their low and
high-energy slopes $\alpha$ and $\beta$. For the peak energy, we assume
a log-normal distribution with a mean value $E_{p,0}$ and a dispersion
0.3 dex. For the slopes, we adopt the observed distribution by Preece et al. (2000).

With these assumptions, the GRB population is entirely
described by four free parameters : $L_\mathrm{c}$ and
$\theta_\mathrm{c}$ for the luminosity function, $k$ for the comoving
rate and $E_\mathrm{p,0}$ for the spectral properties.
 Following the method
 described in DRM06, we constrained them by fitting
 simultaneously: i) the $\log{\rm N}-\log{\rm P}$ distribution of 
 GRBs (where $P$ [ph cm$^{-2}$ s$^{-1}$] is the peak flux)
 detected by the Burst and Transient Source experiment (BATSE; Kommers
 et al. 2000; Stern et al. 2000, 2002); ii) the peak energy
 distribution of bright BATSE bursts (Preece et al. 2000); and iii) the
 HETE2 fraction of X-ray rich GRBs and X-ray flashes (Sakamoto et
 al. ~2005). 
 For a given set of parameters
 $\left(L_\mathrm{c},\theta_\mathrm{c},k,E_\mathrm{p,0}\right)$, a population of $\sim 10^{5}$ 
GRBs is randomly generated, with a redshift $z$, a luminosity $L$ and a
 spectrum characterized by a peak energy $E_\mathrm{p}$ and a low and a
 high-energy slope $\alpha$ and $\beta$. The four free parameters are
 then adjusted to minimize the $\chi^{2}$ obtained when comparing the
 simulated data with the three observations listed above. More details
 about the procedure can be found in DRM06.
 The results for the best fit are $\log(L_{\rm c}[{\rm
erg}\;{s^{-1}}])= 53.7\pm 0.6 $ and $\theta_c = 9.2 \pm 5.2^{\circ}$ for
the luminosity function, $\log({\rm k}) = -5.99 \pm 0.06$ for the
comoving rate and $\log{\left(E_\mathrm{p,0}\mathrm{[keV]}\right)}=2.8\pm 0.1$ for the spectrum.
 The reduced $\chi^{2}= 1.53$
 for 37 degrees of freedom. We note here that DRM06, assuming a power-law luminosity function,
 found $\sim -1.6$ as the best fit value for the slope when this is free
 to vary, as opposed to the slope
 $\sim -2$ predicted by the USJ model (eq.~\ref{eq:lumfun}). However, an acceptable $\chi^{2}$ is also obtained in our
 case.
\begin{figure}
\psfig{file=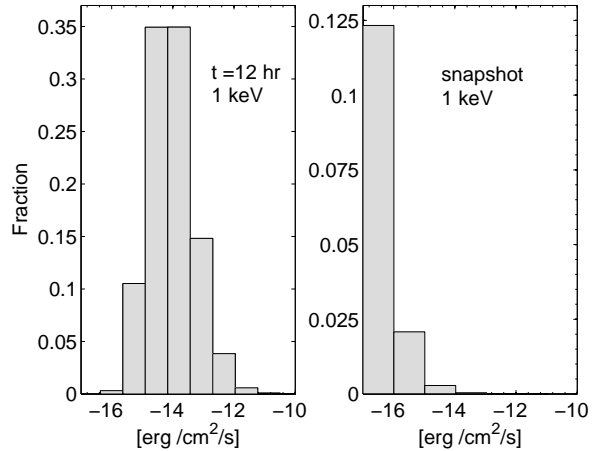,width=0.48\textwidth}
\caption[Intrinsic X-ray flux distribution] {Flux distribution in the X-ray
band at 1 keV energy; these distributions are intrinsic, i.e. no selection criteria have been applied.
We choose the minimum flux, so that the arbitrary cut-off at $10$ yrs for the age 
of an afterglow in our simulations does not affect the shown distributions. 
{\em Left panel}: the
distribution of fluxes at 12 hours after the trigger. This may be
compared with fig.~5 of Berger et al. (2005), taking into account that
their observed distribution suffers from selection effects.{\em
Right panel}: the flux distribution as it appears in a
snapshot observation of the sky.}
\label{fig:xflux}
\end{figure}
\begin{figure}
\psfig{file=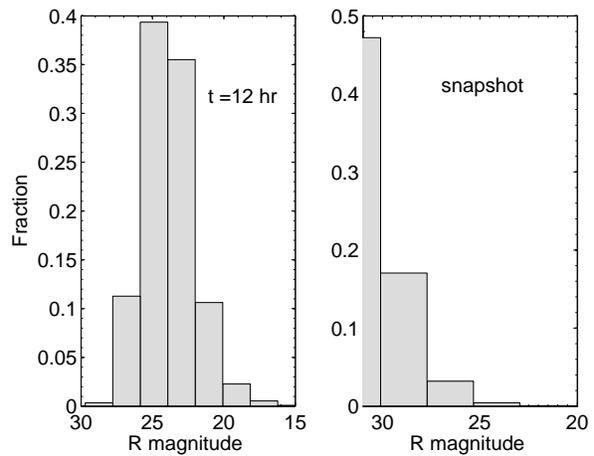,width=0.48\textwidth}
\caption[Intrinsic R magnitude  distribution]
{As Fig.~\ref{fig:xflux} but for magnitude distributions in the R band. 
Our left panel may be compared with fig.~4 of Berger et al. (2005)}
\label{fig:Rflux}
\end{figure}
\begin{figure}
\psfig{file=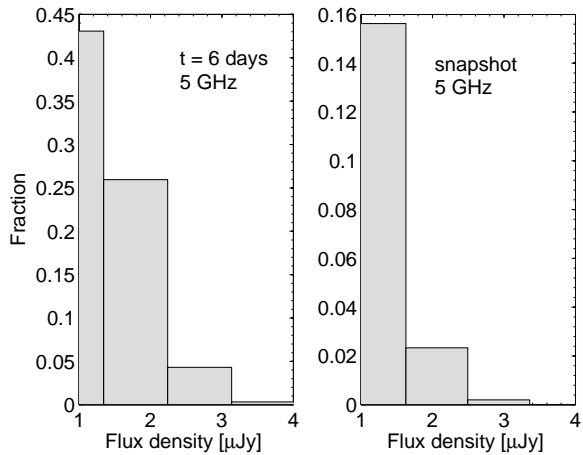,width=0.48\textwidth}
\caption[Intrinsic flux distribution]
{As Fig.~\ref{fig:xflux} but for flux distributions in the radio band at 5~GHz frequency.
Our left panel may be compared with fig.~6 of Berger et al. (2005)}
\label{fig:radioflux}
\end{figure}

\subsection{Physical description of afterglow lightcurves}
\label{sec:code} 
The afterglow emission is modeled as synchrotron radiation from a
relativistic blast-wave propagating in a constant density external medium
(e.g. \Mesz~ \& Rees 1997). We ignore the contribution from the
reverse shock,  although it dominates the afterglow emission in the
first few tens of seconds after the GRB (Sari \& Piran 1999), since OA
observations occur much later. For the same reason,  our
results are largely independent of the choice of the initial Lorentz
factor (which we fix at $\Gamma_{\rm 0}$ = 300), since the deceleration
time is unlikely to exceed $10^3$ s (e.g. Panaitescu \& Kumar
2000). Similar considerations of timing allow us to neglect the modelling of the
early afterglow features (flares, plateau, etc.., e.g. Nousek et al. 2006),
that are not reproduced by the standard external shock model.

 The code we use for the calculation is described in Rossi et
al. (2004). For each lightcurve, $L_{\rm a}$ [erg~s$^{-1}$Hz$^{-1}$],
the input parameters are: the angular distribution of kinetic energy,
the viewing angle, the shock parameters, the external density and the
rest-frame frequency.

The luminosity function parameters, $L_{\rm c}$ and $\theta_{\rm c}$,
constrained by the prompt emission data (\S~\ref{sec:sfr}) allow us
to determine the kinetic energy distribution within the jet
(eq.~\ref{eq:usj}), since

\begin{equation}
 E_{\rm c} \propto \left(\frac{1 - \eta_{\gamma}} {\eta_{\gamma}}\right) L_{\rm c}.
\label{eq:ec}
\end{equation}

\no Inspection of current data suggests that the proportionality
constant is of the order of 1 second.  The $\gamma-$ray efficiency
$\eta_{\gamma}$ that is inferred from modeling data in the standard
external-internal shock scenario is rather high, between $\sim$ 50\%
and $\sim$90\% (Panaitescu \& Kumar 2002). This uncertainty together
with the one $\sigma$ range of $L_{\rm c}$ imply $10^{52} \lsim E_{\rm
c}~ {\rm erg} \lsim 2 \times 10^{54}$. We adopt $E_{\rm c} =1.3
\times 10^{53}$ erg. Since the core angle is not strongly constrained
by prompt emission data (see~\ref{sec:sfr}), we adopt a value at the
low end of the one $\sigma$ range, $\theta_{\rm c} = 4^{\circ}$, to
account for the observed breaks in the lightcurves on day timescales.
The external shock and density parameters are chosen from the ranges
of values inferred from afterglow modeling (e.g. Panaitescu \& Kumar
2001; Panaitescu 2005)\footnote{We are aware that shock and density parameters inferred
from observations are not universal. They rather vary from burst to
burst within some ranges.  To model the emission, however, what is
important is that the combination of these parameters, including the
fireball kinetic energy, gives fluxes comparable
to observations. As mentioned afterwards, we perform such a
comparison. It indicates that we have chosen a reasonable combination
of parameters.}  .  The fraction of energy at the shock that goes into
accelerated electrons and magnetic field is $\epsilon_e = 0.05$ and
$\epsilon_B = 0.005$ respectively; the electrons are accelerated into
a power-law with exponent $p = 2.2$. The external number density is
taken to be $n = 1$ cm$^{-3}$.  The whole set of parameters ($E_{\rm
c}$, $\theta_{\rm c}$, $\epsilon_e$, $\epsilon_B$, $p$ and $n$) yield
results consistent with the observed afterglow flux distributions
(Berger et al. 2005).  Our flux distributions are shown in
Fig.~\ref{fig:xflux} to Fig.~\ref{fig:radioflux}.  In the right
panels, we plot the histogram of fluxes in a given band for all
afterglows detected in a snapshot observation of the sky.  In the left
panels, we show the flux histogram of the same afterglows evaluated at
a common observed time. These latter distributions compare favorably
with fig.~4, fig.~5 and fig.~6 of Berger at al. (2005)\footnote{We do
not attempt a formal quantitative comparison with data, since this
would require us to take into account selection effects in different
bands that are difficult to quantify. }.

\begin{figure*}
\psfig{file=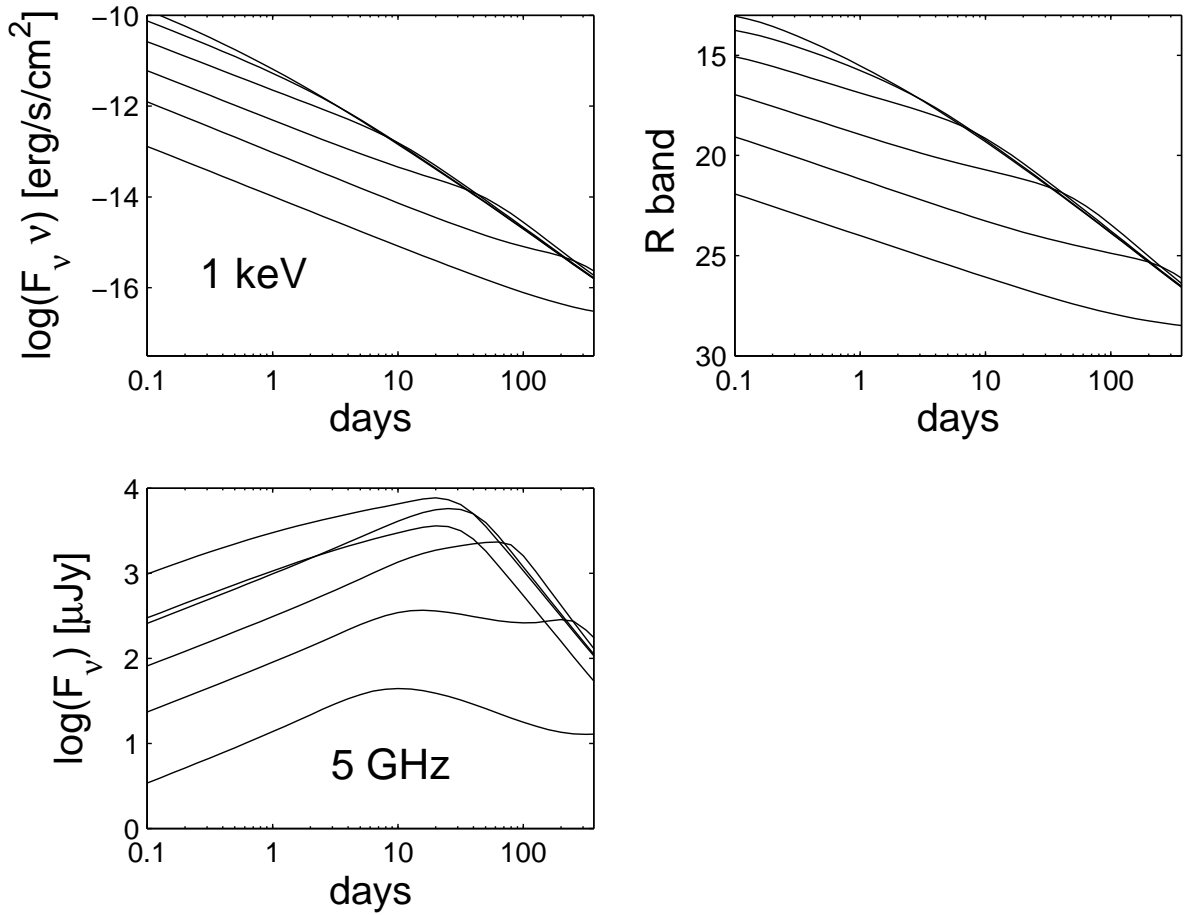,width=\textwidth}
\caption[3 bands lightcurves] {Afterglow lightcurves for a GRB
at z=1 in X-rays (upper left panel), optical (upper right panel) and
radio (lower left panel) bands. For each band, the different curves
correspond to different viewing angles. From top to bottom, $\theta =
0^{\circ}, 4^{\circ}, 8.7^{\circ}, 19^{\circ}, 41.2^{\circ}, 90^{\circ}$.}
\label{fig:lcurves}
\end{figure*}

We compute afterglow lightcurves $L_{\rm a}(\nu \times (1+z), \theta,
t^\prime{})$ for three observed frequencies, $\nu = 2.42 \times
10^{17}$ Hz (1 keV), $\nu = 4\times 10^{14}$ Hz (R band) and $\nu =
5\times 10^{9}$ Hz; 6 redshifts, $0\le $\rm{z}$ \le 20$; 10 viewing
angles, $0^{\circ} \le \theta \le 90^{\circ}$ and 57 comoving times
$t^\prime{}$, spanning 10 years. We arrange those data in the form of a
matrix that can be easily interpolated in order to assign a luminosity
$L_{\rm a}$ to each simulated burst.  The flux observed on Earth is
calculated from $L_{\rm a}$ with $F= L_{\rm a} (1+z)/(4 \pi D_\mathrm{L}^{2})$.
Examples of lightcurves
are shown in Fig.~\ref{fig:lcurves}: in this example the GRB is situated at $z=1$ and
viewed under different angles, in our three bands of observation.


\subsection{Monte Carlo code for "Orphan" Afterglows}
\label{sec:detection}

OAs are generated with a flat probability distribution 
in age, $t_{\rm a}$,
(i.e. the time lag as observed on earth between the GRB explosion and the snapshot observation epoch),
 in an interval of 10 years, at a rate of $R_{\rm obs}\simeq 2195$ yr$^{-1}$
\footnote{This expected rate of GRBs observed from the Earth is given
by integrating $R_{\rm GRB}/(1+z)= k R_{\rm SN} /(1+z)$ over the whole
volume of the universe.}. For the sensitivities of interest in this
paper, afterglows of ten years or older are undetectable. Their mean
duration above threshold is indeed shorter than our total simulated
time (see Figs.~\ref{fig:n&Tx},~\ref{fig:n&TR}
and~\ref{fig:n&Tradio}). Therefore, the arbitrary cut off at ten years
does not influence our results.

We generate each GRB redshift, z, according to the probability distribution
discussed in \S~\ref{sec:sfr} and its viewing angle $\theta$ according
to
\begin{equation}
P(< \theta) = (1 - \cos{\theta}),
\end{equation}
\no with $0^{\circ} \le \theta \le 90^{\circ}$. From the matrix of the
afterglow luminosities interpolated at $t_{\rm a}/(1+z)$, $\theta$ and
$\nu\,(1+z)$, we calculate the OA flux and compare it with a given
flux threshold, $F_{\rm th}$.  We can, thus, compute the expected
number of afterglows above $F_{\rm th}$ in a given band $\nu$: $N_{\rm
snap}(>F_{\rm th},\nu)$.  This is the number that would be seen in a
snapshot observation of the entire sky.

 Model predictions for a flux limited survey require the calculations
of two other quantities.  First, the mean of the time interval $t_{\rm
th}$ spent by an afterglow above the flux limit,

\begin{equation}
 \left<\log_{10}{t_{\rm th}}\right>=\frac{\sum_{0}^{N_{\rm snap}} \log_{10}{t_{\rm th}}}
{N_{\rm snap}} \; \equiv \log_{10}{T_{\rm th}},
\label{eq:tth} 
\end{equation}

\no 
where we actually compute the geometric mean of $t_{\rm th}$ to avoid being biased by extreme
values.
Second, the mean rate at which afterglows appear in the sky over the survey flux threshold,

\begin{equation}
R_{\rm oa}  = N_{\rm snap} \left(\frac{\sum_{0}^{N_{\rm snap}} t_{\rm th}^{-1}}
{N_{\rm snap}}\right) \; \equiv \frac{N_{\rm snap}} {T_{\rm rate}}. 
\label{eq:trate} 
\end{equation}

We performed 160 simulations of flux limited all sky snapshots in
radio and optical bands and 400 in X-rays, where we needed more
statistic.  We then computed the average $N_{\rm snap}$, $T_{\rm
rate}$ and $T_{\rm th}$.  The results are shown in
Figs~\ref{fig:n&Tx},~\ref{fig:n&TR} and~\ref{fig:n&Tradio} (thick
lines).  They can be used to estimate the number of OAs expected in a
given survey. If the observing time of a survey $T_{\rm obs}$ is
shorter than the time $T_{\rm th}$ for which an afterglow is
detectable, then we can consider it a snapshot observation and the
total expected number of detected OAs is \be N_{\rm oa}(> F_{\rm
th},\nu) = N_{\rm snap} (>F_{\rm th}, \nu) \,\,\frac{\Omega_{\rm
obs}}{4\,\pi},
\label{eq:oa1}
\ee

\no
where $\Omega_{\rm obs}$ is the solid-angle of the sky covered by the snapshot.
Vice versa, the total expected number of OAs is computed as
\be
N_{\rm oa}(> F_{\rm th}, \nu) = R_{\rm oa}(F_{\rm th},\nu)~T_{\rm obs} \,\,\frac{\Omega_{\rm obs}}{4\,\pi}.
\label{eq:oa2}
\ee
\no 


\section{Results}
\label{sec:results}

Our results show that, as expected, the probability of detection and
the mean duration above threshold increases with the survey
sensitivity (Figs.~\ref{fig:n&Tx},~\ref{fig:n&TR}
and~\ref{fig:n&Tradio}).  A chance of detection in an all sky snapshot
($N_{\rm snap} \gsim 10$) requires: a flux limit of $\nu
F_{\nu} \lsim 10^{-14}$ erg cm$^{-2}$ s$^{-1}$ at 1 keV ;  a limit
magnitude of $R \gsim 23$ in R band and  a flux density threshold of
$F_{\nu} \lsim 1$ mJy at 5 GHz.

\subsection{Specific surveys predictions}
\label{sec:surveys}

In this section, we provide illustrative predictions for various
current and planned surveys. We use here the results shown in
Figs~\ref{fig:n&Tx}, ~\ref{fig:n&TR} and~\ref{fig:n&Tradio} and eq.~\ref{eq:oa1} and eq.~\ref{eq:oa2}.
 We note
that the following numbers should be taken as upper limits when compared with
observations, since we did not consider detection limitations
(e.g. host galaxy, dust absorption etc..) other than instrumental.
 
We also note that a detailed comparison between
observation and theory requires knowledge of the specific survey
strategies. However, generic conclusions can be drawn from the following examples.

\subsection{X-rays}
\begin{figure*}
\begin{center}
\psfig{file=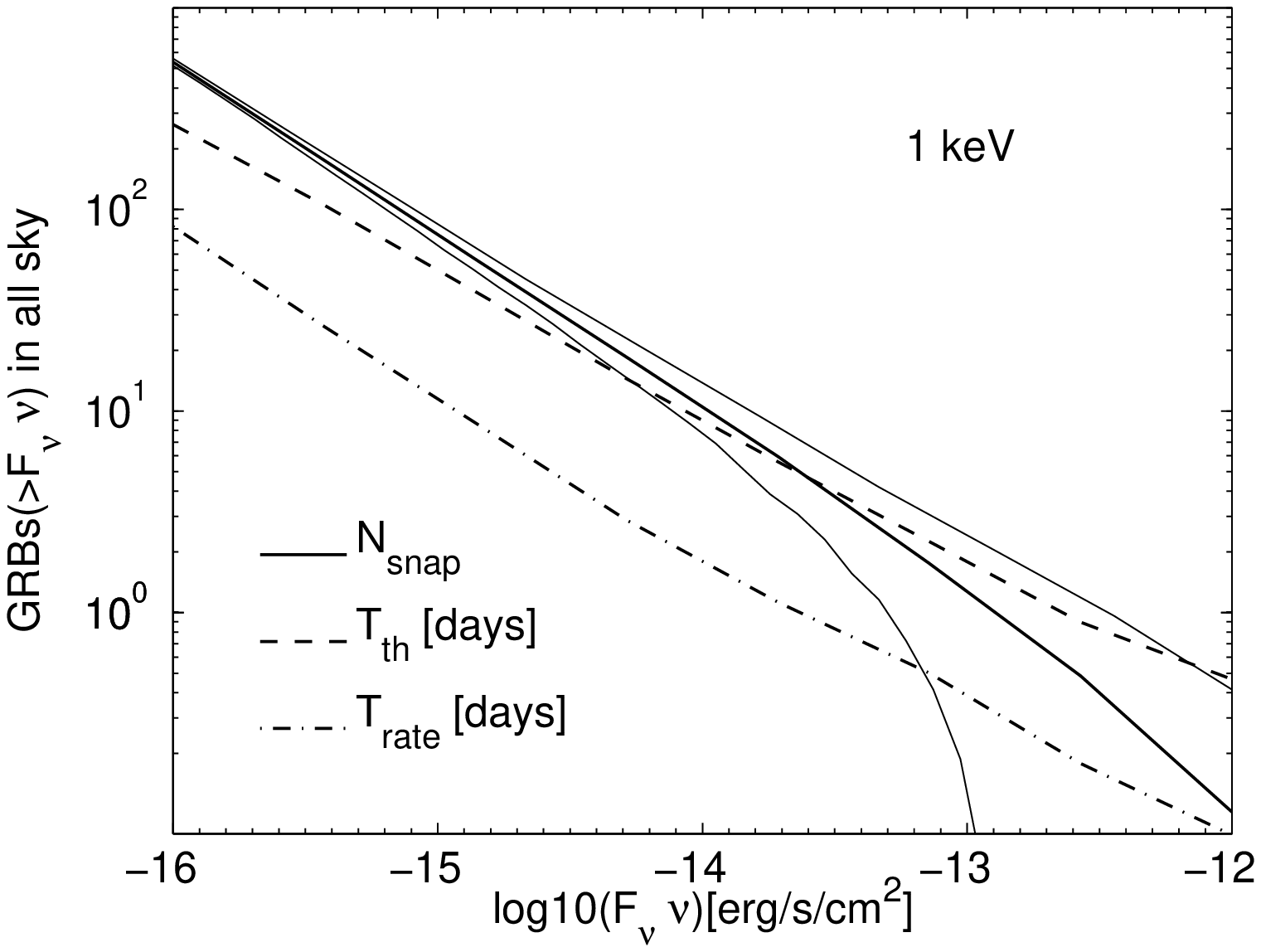,width=\textwidth}
\caption[$N_{\rm shot}$ and $T_{\rm th}$ in X-rays] {Expected number
 of afterglows in an all-sky snapshot observations in the X-ray band,
 as a function of the flux threshold (solid thick line). The thin solid lines are the $1\sigma$ contours.
 We also plot $T_{\rm th}$
 (thick dashed line) and $T_{\rm rate}$ (thick dot-dashed line) in days.}
\label{fig:n&Tx}
\end{center}
\end{figure*}
The two most sensitive X-ray surveys ($ \sim 1 \times 10^{-15}$ erg
s$^{-1}$ cm$^{-2}$ in the 0.5-2 keV band) performed by {\it Chandra}
 and {\it XMM-Newton} are the 2~Msec
Chandra Deep Field-North and the 0.8~Msec XMM-Newton Lockman Hole
field. They cover respectively an area of 0.13 and 0.43
deg$^{2}$. Since $T_{\rm th} \simeq 52$ days, these surveys are
equivalent to two snapshot observations.  The small coverage of the
sky results in a very small detection probability of a few
$10^{-4}$. Larger surveys, as the XMM-Newton Bright Serendipitous
Source Sample (Della Ceca et al. 2004) or The Chandra Multi-wavelength
Project (ChaMP; e.g. Kim et al. 2007), are less sensitive
($ \nu F_{\rm th} \gsim 10^{-14}$ erg s$^{-1}$ cm$^{-2}$) and the
increase in area is not sufficient to yield more than $N_{\rm oa} \lsim 10^{-2}$.

The ROSAT All Sky Survey (RASS; e.g. Voges et al. 1999) covers the
full sky. The RASS exposure is 76435 deg$^{2}$ days, with a
sensitivity of $10^{-12}$ erg s$^{-1}$ cm$^{-2}$.  Therefore the
survey is equivalent to an all sky observation with $T_{\rm obs} =
1.85$ days.  At this flux threshold, OAs are fast X-ray transients,
lasting for $T_{\rm th} \simeq 0.5$ days and $R_{\rm oa} \simeq 0.1$
day$^{-1}$.  Therefore, despite the large coverage of the sky, we
predict that the survey should have found only $N_{\rm oa} \simeq 0.2
$. Greiner et al. (2002) found 23 OA candidates. After
spectroscopic follow-up, however, they concluded that most, if not all
events, are stellar flares. This is in agreement with our predictions.

The prospects for detection are not exciting even for the future
mission eROSITA (extended ROentgen Survey with an Imaging Telescope
Array).  It will perform the first imaging all-sky survey up to 10 keV
with a sensitivity of $5.7 \times 10^{-14}$ ergs$^{-1}$ cm$^{-2}$ in
the 0.5-2 keV band. We predict $\sim 2.3$ OAs.
 
 These dispiriting results can be understood from Fig.~\ref{fig:xflux}
 and Fig.~\ref{fig:n&Tx}: the flux limit should be $\lsim 10^{-14}$
 erg s$^{-1}$ cm$^{-2}$ in order to have a good chance of detection
 in one all sky snapshot. To achieve such a sensitivity, a long
 exposure time is needed (e.g. $\sim$ 50 ksec for Chandra). Given the
 small field of view of the current instruments, a full sky scan is
 unfeasible. We conclude that X-ray surveys are a poor tool for OA
 searches if jets are described by the USJ model.

 \subsection{Optical}
\begin{figure*}
\begin{center}
\psfig{file=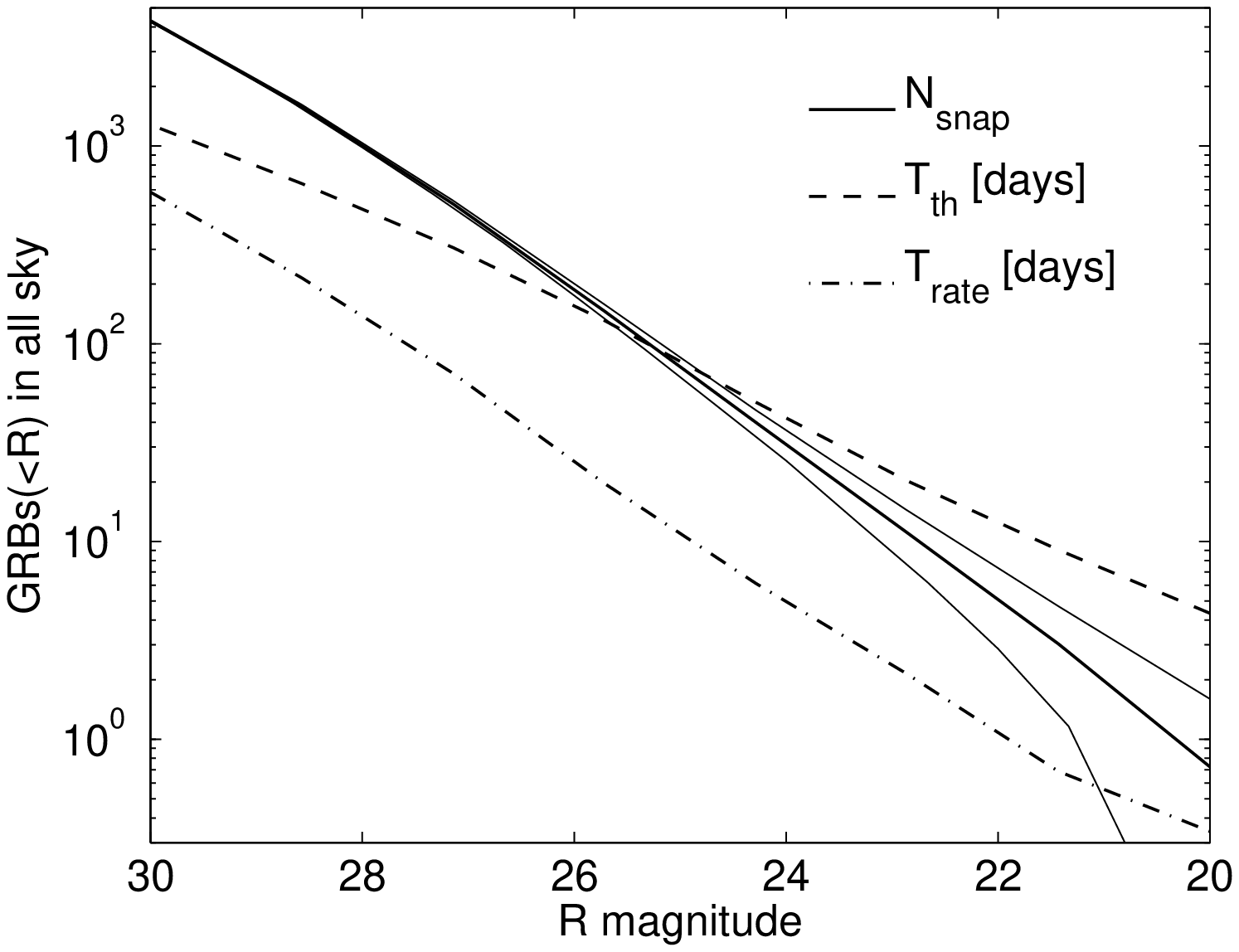,width=\textwidth}
\caption[$N_{\rm shot}$ and $T_{\rm th}$ in R band ] {The same as
Fig.~\ref{fig:n&Tx}, but for the R band.}
\label{fig:n&TR}
\end{center}
\end{figure*}

   The most recent optical OA searches are the ones by Rau et
  al. (2006) and by Malacrino et al. (2007).  Rau et al. observed 12
  deg$^{2}$ of sky for 25 nights, separated by one or two nights. They
  used the MPI/ESO Telescope at La Silla, reaching R=23. Since at that
  sensitivity $T_{\rm th} \simeq 22$ days and $T_{\rm rate} \simeq 2$
  days, we compute the expected OAs through an average rate of $R_{\rm
  oa} = 6.3$ day$^{-1}$.  We find $N_{\rm oa} = 0.02$, considering an
  actual observing time of 12.5 days.  Malacrino et al. used images
  from the CFHTLS very wide survey . They searched 490 deg$^{2}$ down
  to R=22.5.  They observed 25-30 deg$^{2}$ every month over a period
  of 2-3 nights (Malacrino et al. 2006). Since $T_{\rm th} \simeq 17$
  days, we can consider their survey a snapshot observation and we
  predict $N_{\rm oa} = 0.1$. Recently, Malacrino et al. (2007b) have
  rejected the only candidate they had identified in their first work.
  
  The most recent Sloan Digital Sky Survey (SDSS) data release
  (Adelman-McCarthy et al. 2007) includes imaging of 9583 deg$^{2}$,
  with a R magnitude limit of 22.2. We expect $N_{\rm oa} \simeq 1.5$.
  If we take a more conservative magnitude limit of 19 to account for
  the need for spectroscopic identification, the detection probability
  drops to $\sim 0.05$.
  
  The previous meagre results are due to the fact that a very large
  detection area is essential for these sensitivities
  (fig.~\ref{fig:Rflux} right panel and fig.~\ref{fig:n&TR}).  Only
  with the flux limit of the Subaru Prime Focus Camera (R$\simeq 26$
  in 10 minutes of exposure), we could restrict ourselves to 5 \% of
  the sky and get a snapshot with $N_{\rm oa} \simeq 10$.  This,
  however, would require a total observing time of $T_{\rm obs}\simeq$
  57 days.

 Future larger surveys include GAIA (Parryman et al. 2000) and the
Panoramic Survey Telescope \& Rapid Response System (Pan-Starrs;
e.g. Kaiser et al. 2002). The first is an all sky survey with a
magnitude limit of $R \sim 20$. It will observe each part of the sky
60 times separated by one month (Lattanzi et al. 2000). At this
sensitivity, an OA stays in the sky on average for $\sim 3.8$ days; thus GAIA
will perform 60 independent snapshots of the sky, with a prediction of
$N_{\rm oa}\simeq 44$.  Pan-STARRS is expected to scan three-quarters
of the entire sky in about a week, down to an apparent magnitude of
24. Since $T_{\rm th} \simeq 43$ days, this can be considered a
snapshot observation and we expect $\simeq 23$ OAs.  Finally,
the Large Synoptic Survey Telescope (LSST) is planned to cover
10,000 square degrees every three nights down to a depth of R$\simeq 24.5$ (Ivezic et al. 2008).
This would yield $\sim 13$ afterglows every three nights.
Both  Pan-STARR and LSST will be able to repeatedly observe an afterglow 
source, thus monitoring its variability and enhancing the chances of
identification.

Thus, future optical surveys could be powerful tools for OA searches.

\subsection{Radio}
\begin{figure*}
\begin{center}
\psfig{file=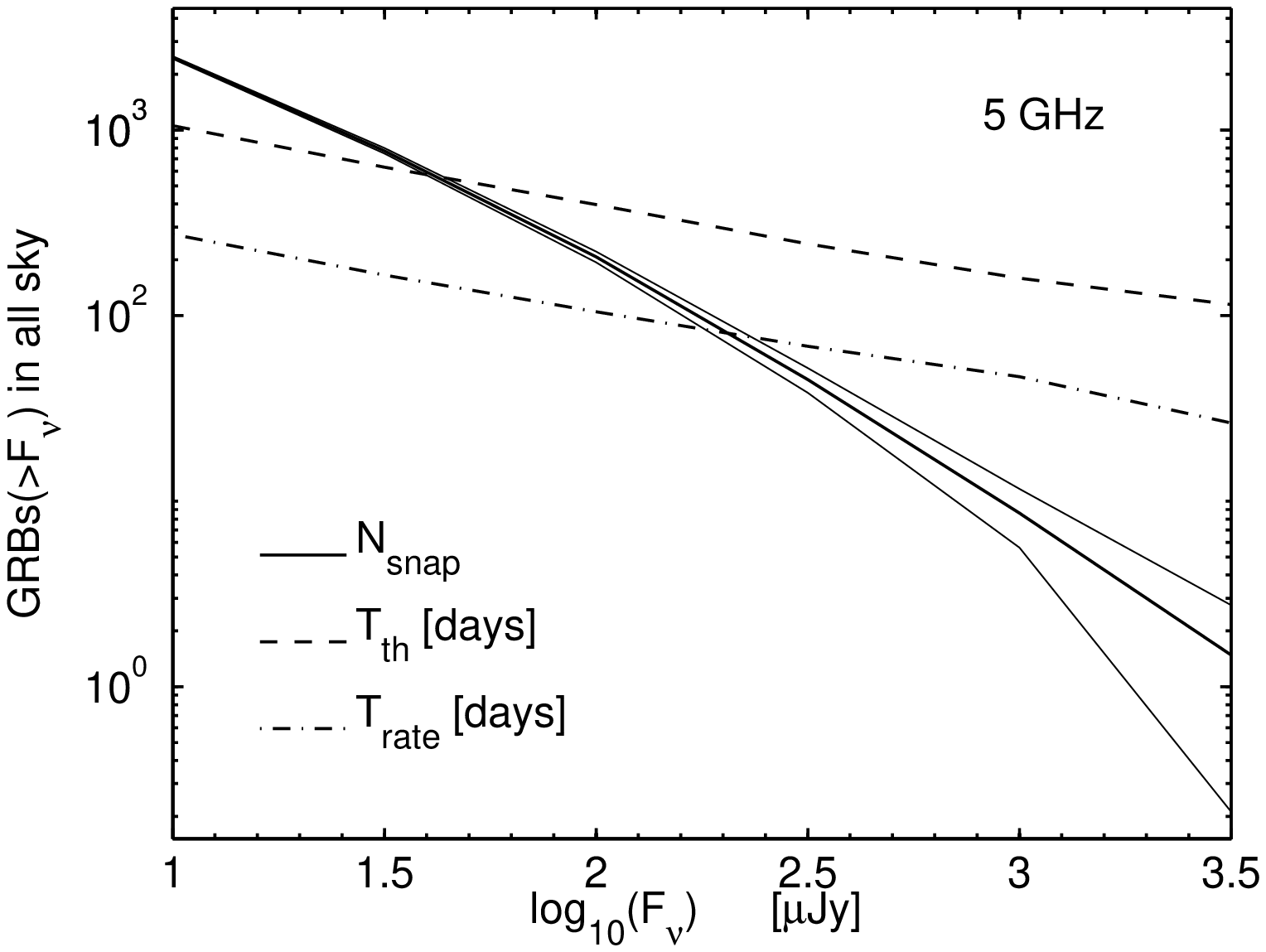,width=\textwidth}
\caption[$N_{\rm shot}$ and $T_{\rm th}$ in radio band] {The same as
Fig.~\ref{fig:n&Tx}, but for the radio band.}
\label{fig:n&Tradio}
\end{center}
\end{figure*}
Levinson et al. (2002) searched for transients by comparing the FIRST
and NVSS (NRAO VLA Sky Survey) radio catalogues and found 9
candidates. Gal-Yam et al. (2005) rejected all candidates by means of
follow-up radio and optical observations and placed an upper limit
($95\%$ confidence) of 65 radio transients for the entire sky above
6~mJy at 1.5~GHz. This may be translated to a sensitivity threshold of
3.3~mJy at 5~GHz, using a typical late time spectral shape in radio of
$F \propto \nu^{-0.5}$ (TP02). We predict $\simeq 1.4$ radio
afterglows.

Recently, Bower et al.(2007) published an archival survey with data
from the Very Large Array, spanning 22 years. For an effective area of
$10~\deg^{2}$, we get a rate of $\simeq 4~\times~10^{-2}$ yr$^{-1}$
for OAs brighter than $370~\mu$Jy. This rate is too low to account for
the 10 detected transients.  In addition the observed transient
duration of approximately a week suggests that those sources are not
indeed afterglows, which are expected to last above that threshold for
approximately half a year.

Fig.~\ref{fig:radioflux} and Fig.~\ref{fig:n&Tradio} and the above
examples show that the OA search would greatly benefit from lowering
the survey sensitivity below 1~mJy, with an area of $\gsim 10,000
\deg^{2}$.

FIRST (Faint Images of the Radio Sky at Twenty-cm; Becker, White \&
Helfand 1995) has covered over $10^{4} \deg^{2}$ of the North Galactic
Cap.  The survey area has been chosen to coincide with that of the
SDSS. The sensitivity is $F_{\rm th} \sim 1$ mJy at 1.4~GHz.  This may
be translated into a flux limit of 0.5 mJy at 5~GHz. If we consider,
as for the SDSS, an area of 9583 $\deg^{2}$, we expect $\simeq 7$ OAs.

The plan is for the Allen Telescope Array (ATA)\footnote{See
e.g. http://ral.berkeley.edu/ata/science/.} to observe $10^{4}
\deg^{2}$ at mJy sensitivity and later to go as deep as $\sim 0.1$~mJy
at 5GHz. The expected number of OAs would then rise from a few to 50.

\section{comparison with the ``top-hat'' model}
\label{sec:USJvsUJ}
\begin{figure*}
\begin{center}
\psfig{file=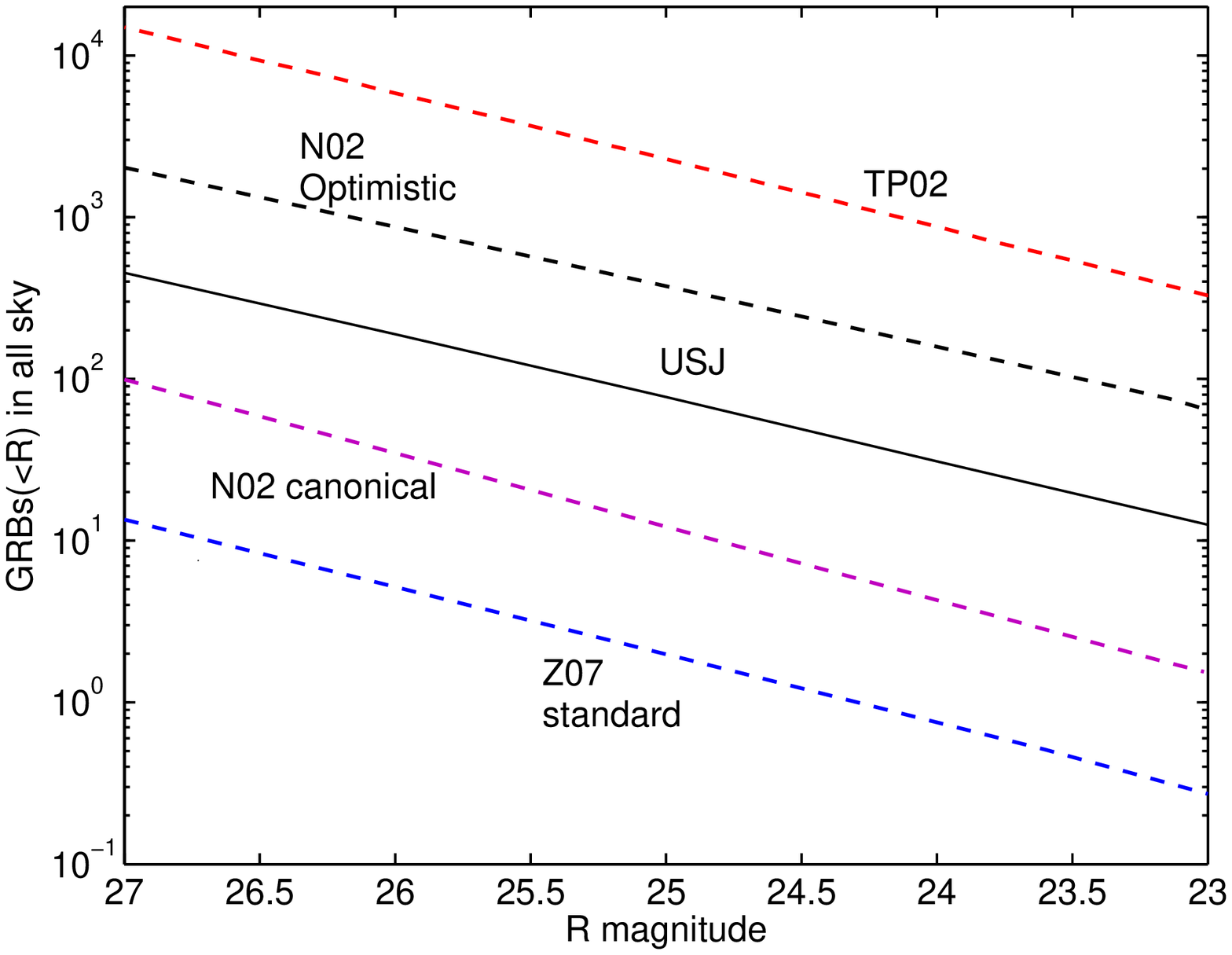,width=\textwidth}
\caption[Comparison with TH] {Comparison between the predicted number of OAs in the TH and USJ model in R band.
From bottom to top, we plot for the TH model the ``standard'' model by Zou et al. (2007); the preferred model and the
optimistic model by Nakar et al. (2002) and the model by TP02. The solid line is the prediction of the USJ model.}
\label{fig:thcomp}
\end{center}
\end{figure*}

In fig.~\ref{fig:thcomp}, we compare our results for $N_{\rm snap}$  (solid line)
with published predictions for the TH model in R-band (dashed lines).
From bottom to top, we plot the ``standard`` model by Z07; the ``preferred`` and the
``optimistic`` models by N02 and the model by TP02.
The curves from these works have similar slopes but they vary in normalisation by several orders of magnitude. 
The main difference is the assumed afterglow luminosity function.

This is the result of different choices for the
opening angle and total jet energy distributions and the afterglow radiation efficiency.
Z07 assume constant total (beamed corrected) energy $E_{tot}=10^{51}$ erg and 
a power-law distribution of opening angles $P(\theta_{\rm jet}) \propto \theta_{\rm jet}^{-1}$.
N02 assume constant peak flux at $\theta_{\rm obs}=\theta_{\rm jet}$
 and a fixed average opening angle $\theta_{\rm jet}=0.1~ {\rm rad}$
(``canonical'' model) and $\theta_{\rm jet}=0.05~ {\rm rad}$ (``optimistic'' model).
Finally, TP02 assume that the whole GRB population is represented by 10 well-studied 
events. We also note that within each model there is an order of magnitude uncertainty.
For example, Zou et al. results differ by an order of magnitude between $E_{\rm tot} =5 \times 10^{50}$ erg and
$E_{\rm tot} =5 \times 10^{51}$ erg (their fig.~4).

These uncertainties in the TH model are due to the fact that
neither the  opening angle distribution nor the total jet energy distribution 
is uniquely defined by the model or unbiasely determined by observations 
(e.g. through the observed luminosity function).
The USJ model, instead, has more predictive power, since it has less degrees of freedom.

From the comparison in fig.~\ref{fig:thcomp}, one concludes that our model predictions fall roughly in the middle
of the range of published results for the TH model. The differences in the predictions between our model and 
any {\em individual} TH model are large enough to be significant, but there is a sufficiently wide class 
of plausible TH models to encompass almost any luminosity function of OA that might be observed 
in the future. In principle, therefore, OA observations could rule out the USJ model 
(noting of course that there is a similar level of uncertainty to our predictions as individual TH models), but the same 
cannot be said for the TH model. The TH model could be tested by a combination of orphan afterglow 
observations and independent constraints on the parameters of the model.


\section{Discussion and Conclusions}
\label{sec:fine}

The realization that GRBs may be jetted has triggered studies of
orphan afterglows. Early studies assumed that the prompt GRB
emission comes from a sharp jet, in which case the ratio of afterglows to that of
GRBs yields a constraint on the GRB beaming fraction
(or equivalently the opening angle of the GRB jet). This parameter is of
importance for a proper assessment of GRB rates and energetics.

The structured jet model has offered an equivalent explanation for
afterglow phenomenology. However, its interpretation of the lightcurve
breaks is different: they would arise from viewing angle effects and
not from geometrical collimation.  In fact, if GRB jets are indeed
structured, most, if not all, afterglows should be generally preceded by a prompt
emission pointing towards the observer. Therefore, even if in practice
the relative number of detections in various bands depends on the
survey strategy, the ratio should tend to unity ($b=1$) 
if events are detectable at arbitrarily low fluxes.

 In this paper, we have investigated the detection prospects of afterglows
 for flux limited surveys, in the USJ framework.
 We conclude that large sky coverage is
 essential in all bands. In addition, X-ray and radio instruments
 should push their flux limit below $10^{-14}$ erg s$^{-1}$
 cm$^{-2}$ (at 1 keV) and $\sim 1$ mJy (at 5 GHz) respectively.
 Current and planned X-ray surveys are thus not suited for OAs searches,
 if the jet is structured.
 The FIRST and the future ATA projects could be successful in detecting radio OAs.
 The potential is even better for future optical all-sky
 surveys, such as GAIA and Pan-Starrs. We also note that it would be worthwhile to exploit the great
 sensitivity of the Suprime-Cam, for which 5-10\% of the whole sky
 is sufficient for positive detections. Certainly, a combination of
 X-ray, optical and radio observations would yield the most of information
 and will help understanding whether the USJ describes correctly the structure of the jet.


\section*{Acknowledgments}
We are very grateful to T.~Totani and A.~Panaitescu for providing us with their data
for the TH predictions. We also acknowledge very useful discussions
 with G. Bower and E. Nakar.  
 EMR acknowledges support from NASA though Chandra Postdoctoral Fellowship 
grant number PF5-60040 awarded by the Chandra X-ray Center, which is operated 
by the Smithsonian Astrophysical Observatory for NASA under contract NASA8-03060.

\end{document}